\documentstyle[11pt,epsf]{article}

\def\beq{\begin{eqnarray}}
\def\eeq{\end{eqnarray}}
\def\bea{\begin{eqnarray*}}
\def\eea{\end{eqnarray*}}

\def\centeron#1#2{{\setbox0=\hbox{#1}\setbox1=\hbox{#2}\ifdim
\wd1>\wd0\kern.5\wd1\kern-.5\wd0\fi
\copy0\kern-.5\wd0\kern-.5\wd1\copy1\ifdim\wd0>\wd1
\kern.5\wd0\kern-.5\wd1\fi}}
\def\ltap{\;\centeron{\raise.35ex\hbox{$<$}}{\lower.65ex\hbox{$\sim$}}\;}
\def\gtap{\;\centeron{\raise.35ex\hbox{$>$}}{\lower.65ex\hbox{$\sim$}}\;}

\def\singleandthirdspaced{\baselineskip=\normalbaselineskip\multiply
    \baselineskip by 130\divide\baselineskip by 100}
\def\singlespaced{\baselineskip=\normalbaselineskip}

\newcommand{\newc}{\newcommand}
\newc{\qbar}{{\overline q}}
\newc{\Kahler}{K\"ahler }
\newc{\deltaGS}{\delta_{\rm GS}}
\begin{document}
\begin{titlepage}
\begin{flushright}
{\large hep-th/9909020 \\ SCIPP-99/22\\

PRELIMINARY}
\end{flushright}

\vskip 1.2cm

\begin{center}

{\LARGE\bf Non-Renormalization Theorems for Operators with
Arbitrary Numbers of Derivatives in
${\cal N}=4$ Yang Mills Theory}

\vskip 1.4cm

{\large Michael Dine and Josh Gray}
\\
\vskip 0.4cm
{\it Santa Cruz Institute for Particle Physics,
     Santa Cruz CA 95064  } \\

\vskip 4pt

\vskip 1.5cm

\begin{abstract}

We generalize the proof of the non-renormalization of the
four derivative operators in ${\cal N}=4$ Yang Mills theory with
gauge group $SU(2)$
to show that certain terms with $2N$ derivatives
are not renormalized in the theory with gauge group $SU(N)$.
These terms may be determined exactly by a simple perturbative
computation.  Similar results hold
for finite ${\cal N}=2$
theories.  We comment on the implications of these results.

\end{abstract}

\end{center}

\vskip 1.0 cm

\end{titlepage}
\setcounter{footnote}{0} \setcounter{page}{2}
\setcounter{section}{0} \setcounter{subsection}{0}
\setcounter{subsubsection}{0}

\singleandthirdspaced


\section{Introduction}

${\cal N}=4$ Yang-Mills theory is a remarkable theory in a number of
ways.  It is a finite and scale invariant.  It is believed to
exhibit an exact electric-magnetic duality.  It also plays a
crucial role in the Matrix model and in the AdS/CFT duality.

There is evidence that the quantum properties of this theory are
even more remarkable than just finiteness.  It has been shown, for
example, that not only are the terms with two derivatives not
renormalized, but the four derivative terms are not renormalized
as well\cite{dsnr}.
The first suggestion of such a possibility was provided by the
agreement of graviton-graviton scattering with the Matrix
model\cite{bfss}.  Subsequently, it has been observed that three
graviton\cite{yoneya} and even $N$-graviton scattering\cite{deg} at
tree level in
supergravity agree with the predictions of the matrix model.   While
some of these analyses are specific to the case of the matrix model
in $0+1$ dimensions (corresponding to non-compact eleven-dimensional
space), some of these hold in higher dimensions, including
$3+1$\cite{deg}.
This suggests that there should be non-renormalization theorems
for $2N$ derivative terms in Yang-Mills theory with gauge group
$SU(N)$ in various dimensions up to four.   In $0+1$ dimensions,
such theorems have been proven for terms with four\cite{pss}
and six derivatives\cite{sethistern}.
In four dimensions, only the four derivative terms have been
shown to be unrenormalized.

In the present note, we generalize the arguments of \cite{dsnr} to
show that certain $2N$ derivative terms are not renormalized in $4$-dimensional
${\cal N}=4$ $SU(N)$ Yang Mills theory.  The
strategy is similar to that used to calculate multigraviton
scattering amplitudes in the matrix model in \cite{deg}.  The
theory has a large moduli space; at generic points, the gauge
symmetry is $U(1)^{N-1}$.  One can, however, consider regions of
the moduli space in which there is a hierarchy of breakings;
$SU(N)$ is broken to $SU(N-1) \times U(1)$, then to $SU(N-2)\times
U(1) \times U(1)$, and so on.  At each stage of this breaking, one
can integrate out the most massive fields and obtain a suitable
effective lagrangian.  By focusing judiciously on certain terms in
these effective lagrangians, one can make arguments similar in
spirit to those of \cite{dsnr}.

In the rest of the paper, we present the proof.  In the next
section, we review the analysis of \cite{deg}
with particular emphasis on the case of $3+1$ dimensions
(corresponding to compactifying three of the $M$-theory
dimensions).  In section three, we generalize the argument of \cite{dsnr} for
non-renormalization of $F_{\mu \nu}^4$ and other four
derivative terms in $SU(2)$ to a
statement about {\it certain} such operators in $SU(N)$ (while we
suspect, as argued in \cite{lowe}, the statement holds in
general, we will not attempt to prove it).  In section 4, we
show that certain six derivative terms in $SU(3)$ are not renormalized.  The
strategy is to first look at an effective $SU(2)$ symmetric theory,
and represent the effect of integrating out the heavy fields through
a suitable spurion.  The symmetries -- scale invariance and $U(1)_R$
invariance, and an approximate shift symmetry for
the background dilaton multiplet -- are sufficient to completely determine certain six
derivative terms in the theory.  This argument can be generalized
to $SU(N)$; this is presented in section 4.  In section 5, we note
that identical arguments and results hold for the finite ${\cal N}=2$
theories.
Section 6 contains
some speculations.

\section{Graviton scattering in $8$ Dimensions and It's
Implications for ${\cal N}=4$ Yang Mills Theory}

The principle reason to suspect that there exists a large
hierarchy of non-renormalization theorems comes from studies of
multigraviton scattering in the matrix model.  The agreement found
in three graviton scattering in $11$-dimensional Minkowski space \cite{yoneya}
is impressive, and suggests that there should be
non-renormalization theorems for some set of six derivative terms
in the Matrix quantum mechanics.  In \cite{deg}, it was shown that
there is actually agreement for certain terms in $N$-graviton
scattering, for arbitrary
$N$.  Moreover, this agreement persists when the theory is
compactified on tori of 1,2 or 3 dimensions.  As a result, one
expects an infinite set of non-renormalization theorems in these
theories.  The strategy of the proof will be closely related to
the approach of these earlier computations, so it is perhaps
useful to review them here.  We will consider specifically the
case of compactification of $3$ dimensions on a small torus,
corresponding to ${\cal N}=4$ Yang-Mills theory on a large
torus\cite{tayloretal,susskindetal}.

To study $N$-graviton scattering (in the Discrete Light
Cone (DLCQ) formulation of the theory) in the theory
compactified to $8$ dimensions, one considers the ${\cal N}=4$ Yang-Mills
theory with gauge group $SU(N)$.  This theory has a moduli space;
at generic points the symmetry $SU(N)$ is broken to $U(1)^{N-1}$,
with the moduli of this breaking being identified with the
coordinates of $N$ gravitons.  If we write the ${\cal N}=4$ theory in
terms of six real (matrix-valued) scalar fields, $\phi^i$, then
\beq
\phi^i = \left ( \matrix{v^i_1 + \phi^i_1 & 0 &0 & \dots \cr 0 & v^i_2 +
\phi^i_2 & 0 &\dots
\cr 0 & 0 & v^i_3 + \phi^i_3 & \dots \cr \dots & \dots & \dots & \dots } \right )
\eeq
It will be notationally convenient to extend the group to $U(N)$
so that the $v_i$'s are unconstrained.

The approach of \cite{deg} was to consider a hierarchy of
expectation values, $v_N \gg v_{N-1} \gg \dots v_2 \gg v_{1}$.
One can then
think of a sequence of breakings, first to $SU(N-1) \times U(1)$,
then to $SU(N-2) \times U(1) \times U(1)$, and so on.  To
illustrate the procedure, consider first the case of $SU(3)$.  In
this case, the large expectation value
of $\phi_1$ breaks the symmetry first to
$SU(2) \times U(1)$ ($\times U(1)$, for $U(3)$).
We can consider the
effective lagrangian for the $SU(2)$ theory, obtained by
integrating out the massive states associated with the first stage
of breaking.  At one loop, one can read off the result by a simple
trick, generalizing the $SU(2)$ result (e.g.
\cite{bc}).  This gives ($F_{12} = F_{11}-F_{22}$, etc., i.e. they
are the differences of the diagonal matrix elements)
\beq
{\cal L}= {1 \over 16 \pi^2} \left ( {[(F_{12}^{\mu \nu})^4 + \dots] \over
\vert \phi_{12} \vert^4} +{[(F_{13}^{\mu \nu})^4
+ \dots]  \over \vert \phi_{13} \vert^4}+
{[(F_{23}^{\mu \nu})^4 + \dots] \over \vert \phi_{23} \vert^4} \right ).
\label{firsteffectivel}
\eeq
Here the dots inside the braces denote
\beq
[(F^{\mu \nu})^4 + \dots]= (F_{\mu \nu})^4 - {1 \over 4}
(F_{\mu \nu}^2)^2 +
2 \partial^{\mu} \phi^i \partial_{\mu} \phi^j \partial^{\nu}\phi^i
\partial_{\nu}\phi^j -
\partial^{\mu} \phi^i \partial_{\mu} \phi^i \partial^{\nu}\phi^j
\partial_{\nu}\phi^j.
\eeq

Now expand the denominators in small
fluctuations about the expectation value, i.e.
replace $\phi_i$ in the denominators by
$v_3\delta_{i3} + \phi_i$, and expand to second order
in $\phi_i$.  Identify $\phi_1- \phi_2 = 2\phi^{[3]}$,
where the label in braces denotes the Cartan generator.
We can generalize this
to an $SU(2)$-invariant expression by replacing $\phi^{[3]}\phi^{[3]}$
by $\phi^a \phi^a$.  This yields:
\beq
{\cal L}= \dots +
{9 \over 8 \pi^2}[(F^{[8]}_{\mu \nu})^4 + \dots] \phi^{ai} \phi^{aj}({\delta_{ij} \over
\vert v_3 \vert^6} - 6{v^i_3 v^j_3 \over \vert v_3 \vert^8}).
\eeq
($F^{\mu \nu[8]} \propto F^{\mu \nu}_{13}+F^{\mu \nu}_{23}$ corresponds
to the generator conventionally called $T^8$ in $SU(3)$).

Now, one can contract $\phi^{ai} \phi^{aj}$.  The propagator,
in the presence of
a background $\phi^{[3]}$, is given by
\beq
\langle \phi^{+i} \phi^{-j} \rangle= {\delta^{ij} \over k^2 + M^2}
+ {4 \partial_{\mu} \phi^{[3]i}\partial^{\mu}
\phi^{[3]j}+ \delta^{ij}(\dots) \over (k^2+M^2)^3}.
\eeq
where $M^2 = 2 g^2 \vert v_{12}\vert^2$.  Integrating over $k$
yields six-derivative terms in the low energy effective theory
which agree precisely with tree level calculations in supergravity
(similar statements hold in $11,10$ and $9$ dimensions).
As a result, we expect that, in $SU(N)$, non-renormalization
theorems hold for certain terms with $2N$ derivatives. In the
following, we will show that this is the case, in regions of the
moduli space where the expectation values are hierarchically
ordered, for operators of the form
\beq
{[(F_{\mu \nu}^{[N-1]})^4+\dots] \over  (v_N)^6 }{(\partial_{\mu}
\phi^{[N-2]} \partial^{\mu}
\phi^{[N-2]})
\over (v_{N-1})^4}{(\partial_{\mu}\phi^{[N-3]} \partial^{\mu}\phi^{[N-3]})
\over (v_{N-2})^4}\dots {(\partial_{\mu}\phi^{[1]} \partial^{\mu}\phi^{[1]})
\over (v_{12})^2}.
\label{nderivative}
\eeq
Here, in $SU(N)$ ($N>3$), we are labeling
the elements of the Cartan
subalgebra, $1,\dots N-1$.

\section{The Non-Renormalization Theorem for the four derivative
terms:  Extension to $SU(N)$}

Let us review, first, the proof of the theorem for the group
$SU(2)$.  Presumably, it would be easy to prove the theorem if one
had a convenient superspace formulation for theories with sixteen
supersymmetries.  Lacking this, it was noted in \cite{dsnr} that
one can exploit an ${\cal N}=2$ subgroup of the full supersymmetry, for
which a full off-shell superspace formulation is available.  It is useful to
first describe the theory in an ${\cal N}=1$ language.  The theory
consists of three chiral multiplets, $\phi_i$, and a gauge
multiplet, $W_{\alpha}$, all in the adjoint representation of the
group.  An $SU(3) \times U(1)_R$ subgroup of the full $SU(4)$
R-symmetry is manifest in this description.  The $\phi_i$'s
transform as a triplet, each with charge $+2/3$ under the $U(1)_R$.

In an ${\cal N}=2$ description, the theory consists of a vector multiplet
and a hypermultiplet.  The vector multiplet consists of the ${\cal N}=1$
vector multiplet and one of the chiral fields, say $\phi_1$.  One
can write
\beq
\psi = \phi + \tilde \theta W + \tilde \theta^2 G
\eeq
(we have dropped the subscript $1$ on $\phi$).  The kinetic term
for the $\psi$'s is (explicitly indicating the adjoint $SU(2)$
index)
\beq
{\cal L}_{vec} = \int d^2 \theta \int d^2 \tilde \theta \psi^a
\psi^a.
\eeq
Our focus in the following will be on the vector multiplets.  We
will study flat directions corresponding to expectation values of
the scalar components of these multiplets (only).  Note that when
these fields get expectation values, the $U(1)_R$ symmetry which
we described above is broken, but another $U(1)$, which we will
call $X$, survives, which is a linear combination of the original
$U(1)$ and an $SU(3)$ generator.  The hypermultiplets have charge
$+4/3$ under this symmetry.

The fact that the kinetic term is an integral over half of
superspace allows one to prove many remarkable properties of the
theory.  Perhaps somewhat more remarkable is that in the ${\cal N}=4$
case, one can prove statements that involve integrals over the
full ${\cal N}=2$ superspace.  Consider the flat direction in which $SU(2)$
is broken to $U(1)$ by the scalar field in the vector
multiplet.  Call the light vector multiplet in this direction
\beq
\psi \sim \psi^a \psi^a
\eeq
We can ask what sorts of terms one can write involving an integral
over the full superspace, which respect the symmetries of the
theory.  Such an integral has the form\cite{roceketal}:
\beq
{\cal L}_{\partial^4} =\int d^8 \theta {\cal H(\psi, \psi^{\dagger})}.
\eeq
The theory is conformally invariant and so
${\cal H}$ must be dimensionless.  It must respect the $U(1)_R$
symmetry, under which $\psi$ transforms by a phase.  These
conditions restrict ${\cal H}$ to the form\cite{roceketal,dsnr}:
\beq
{\cal H}= {1 \over 16 \pi^2}\ln(\psi) \ln(\psi^{\dagger}).
\label{hform}
\eeq
 No scale is necessary in the
logarithm, since the dependence on the
scale would vanish after integration over
$\theta$'s.  In other words, this
expression is scale invariant.
Related to this, the integral of ${\cal H}$ vanishes
under an $R$ transformation, since the integral over a chiral
superfield over the full superspace is zero.  If one now includes
a background dilaton field in a vector multiplet, it is easy to
see that this cannot appear in ${\cal H}$ without spoiling both
the scale and $R$ symmetries.  As a result, the one loop
expression for the four derivative terms in the effective lagrangian
is exact.

This effective lagrangian includes terms with four powers of
$F_{\mu \nu}$, as well as terms with derivatives of scalars and
fermions.  If one compares with component field
computations, one finds complete agreement up to terms which
vanish if one uses the lowest order equations of
motion\cite{periwal}.
  It is also not hard to guess a
generalization to $SU(N)$.  There are now $N-1$ massless fields;
one can write them as differences of diagonal entries of an
$N\times N$ matrix, $\psi_{ij}= \psi_i - \psi_j$.  Then a guess
for a generalization of the $SU(2)$ result, which is symmetric
under permutations as well as scale invariance and $R$ symmetries,
is\cite{lowe,dsunpublished}:
\beq
{\cal H} = {1\over 16 \pi^2}\sum_{i<j} \ln(\psi_{ij}) \ln(\psi_{ij}^{\dagger}).
\eeq
This expression respects all of the symmetries.  It agrees with
an explicit one loop computation. If this term were unique,
one could again immediately prove a non-renormalization theorem.
However, the symmetries we have used
up to now do not suffice to uniquely determine ${\cal H}$.
Ratios of different $\psi_{ij}$'s are both scale
invariant and $U(1)_R$ invariant.
In other words, functions such as
\beq
f(\tau,\tau^{\dagger}) {\psi_{ij} \over \psi_{kl}}
{\psi_{mn}^{ \dagger}\over \psi_{op}^{\dagger}}
\eeq
for various choices of $i,j \dots$ are invariant
under the $U(1)_R$ invariance and scale invariance for
any choice of $f$.
It is possible that one can
still constrain the function completely using the full $SU(4)$
R-symmetry, which is not manifest in the ${\cal N}=2$ setup.  We will not
attempt this here.  Instead, we will content ourselves with a more
limited statement about four derivative terms in these theories.

Consider, first, the case of $SU(3)$.  Suppose that one eigenvalue
of $\phi$, say $\phi_3$ is much larger than the others; more
precisely, $\phi_{13} \approx \phi_{23} \gg \phi_{12}$.  In this
limit, we can integrate out the fields with mass of order
$\phi_{13}$ to obtain an $SU(2)$-symmetric (Wilsonian) effective
action.  This action, again, can be written as an integral of a
function over the whole superspace.  It must be scale invariant
and $R$-invariant.  In general, again, it can involve ratios of
the $\psi_{ij}$'s.  But certain operators cannot
be generated by such ratios.  In particular, consider
those terms which involve four factors of
$F_{\mu \nu}^{[8]}.$
$F^{[8]}$ couples only to heavy
fields. On dimensional grounds,
these terms are suppressed by at least four
factors of the expectation value $v_{13}$ or $v_{23}$.   Restricting
our attention to terms with precisely four such factors
limits the possible dimensionless ratios which can be relevant.
In general, ${\cal H}$ could involve $\psi_{12} \over \psi_{13}$,
for example.  But $\psi_{12}$ would contribute either a factor of
$F_{12}$, which is not relevant here, or a factor of $v_{12}$,
which would imply a suppression by a power of $v_{13}$.
Similarly, ${\psi_{13} \over \psi_{23}}=
1+ {\psi_{12} \over \psi_{23}}$ is irrelevant.  As a result,
${\cal H}$ must take the form
\beq
{\cal H}= {1 \over 16 \pi^2}(\ln(\psi_{13})\ln(\psi_{13}^{\dagger}) +
\ln(\psi_{23})\ln(\psi_{23}^{\dagger})).
\eeq
Again, introducing a background dilaton in the theory, one sees
that this coupling is not renormalized.

In this way, we have established that in $SU(3)$, the terms in the
effective action proportional to
\beq
{[(F^{\mu \nu}_{13})^4+ \dots] \over \vert v_{13} \vert^4} + {[(F^{\mu \nu}_{23})^4+ \dots]
\over \vert v_{23} \vert^4}
\eeq
are not renormalized.  This result
clearly generalizes to the case where $SU(N)$ is broken to
$SU(N-1)$, to terms involving
\beq
\sum_{i=1}^{N-1} {[(F_{Ni}^{\mu \nu})^4 + \dots]  \over \vert v_{Ni} \vert^4}.
\eeq
As always, we are assuming the existence of a
suitable Wilsonian effective action.

In support of this argument, one can consider the two loop
corrections to the four derivative terms.  If one examines the
various two loop diagrams, it is easy to see that, in $SU(N)$, in all
of the diagrams, the
term proportional to
$(F^{[N-1]})^4 \over (v_N)^4$
is
proportional to $N(N+1)$.  As a result, the cancellation in the
case of $SU(2)$ (guaranteed by the theorem of \cite{dsnr}) insures
the cancellation to this order in $SU(N)$.

\section{Six Derivative Terms in SU(3)}

First focus on the case of $SU(3)$.  In the flat direction, $\phi$ is a
diagonal matrix.  As in our discussion of the
previous section, we can take it to be principally in the $T^8$
direction, with a small component in the $T^3$ direction, i.e. we
can write  (for simplicity, writing as a $U(3)$ matrix)
\beq
\phi= \left ( \matrix{v_3 & 0 & 0 \cr 0 &  {\phi^{[3]} \over 2} & 0 \cr 0 & 0 &
 - {\phi^{[3]} \over 2}} \right )
\eeq
$v_3$ and $\phi^{[3]}$
are complex.

To study whether the six-derivative operator implied by equation
\ref{nderivative} is renormalized, we might try to use the $N=2$ setup of the
previous section in the full theory.  Six derivative terms would
then correspond to integrals over the full superspace of terms
with four covariant derivatives.  The particular operator would be
generated by terms such as
\beq
\int d^8 \theta f(\tau,\tau^{\dagger})(D_{\alpha} \psi_{12})(D^{\alpha} \psi_{12})
(\bar D_{\dot \alpha} \psi^{\dagger}_{31})(\bar D^{\dot \alpha}
\psi_{31}^{\dagger}){1 \over \vert \psi_{31}
\vert^4 \vert \psi_{12} \vert^2}.
\eeq
This term, one can check, is consistent with all of the
symmetries.  However, it is not easy to show that this is not
renormalized, as for the four derivative terms, since the
integral is scale and $R$-invariant for any choice of the function
$f(\tau,\tau^{\dagger})$ (in particular, it is invariant for
functions $f(\tau-\tau^{\dagger})$, corresponding to possible
perturbative corrections).

Instead, we resort to a different strategy, which closely
parallels the calculation of section 2.
We note that for small $\phi_3$, the low energy theory is
approximately an $SU(2)\times U(1)$ ${\cal N}=4$ supersymmetry gauge
theory.  This theory, for constant background $\phi^{[8]}$
and $F_{\mu \nu}^{[8]}$ possesses not only unbroken supersymmetry
but unbroken $R$ symmetry.  For slowly varying $F_{\mu \nu}^{[8]}$
and $\phi^{[8]}$, supersymmetry is broken,
as is the $U(1)_R$ symmetry.  This breaking is described by
operators which couple these fields to the $SU(2)$ degrees
of freedom.  The leading such operator is obtained from
\beq
{\cal H} = {1 \over 16 \pi^2}\sum_{i=1}^2 \ln(\psi_{3i})\ln(\psi^{\dagger}_{3i}).
\eeq
As before, one expands for small $\phi_1$, $\phi_2$, and obtains
the same $SU(2)$ expression as we did earlier:
\beq
{\cal L}_{eff}={9 \over 8 \pi^2}{[(F^{[8]}_{\mu \nu})^4+ \dots] \over \vert v_3\vert^4}
({ \vert \phi^a \vert^2 \over \vert v_3 \vert
^2}+ {\phi^a \phi^a \over (v_3)^2}+ {\phi^{a\dagger} \phi^{a\dagger} \over
(v_{3}^*)^{2}})
\label{rviolating}
\eeq
The braces now denote
\beq
[(F^{\mu \nu})^4 + \dots] = (F_{\mu \nu})^4 -{1\over 4} (F_{\mu \nu}^2)^2
+ \partial_{\mu}\phi \partial^{\mu}\phi
\partial_{\nu}\phi^{\dagger} \partial^{\nu} \phi^{\dagger}
\eeq
and $\phi$ is a complex
field,  $\phi = \phi^1 + i \phi^2$.

This lagrangian can be viewed as a perturbation of the low energy,
$SU(2)$ theory.  For non-vanishing background
$F^{[8]}$ and $\partial \phi^{[8]}$ it breaks the
supersymmetries.  The last two terms in
eqn. \ref{rviolating} also violate the $U(1)$
R-symmetry, and we will focus on these.
The first point to note about these terms
is that they are not renormalized.  This is
established by our earlier proof of the non-renormalization of the
four derivative terms
\footnote{One might object that
in our earlier non-renormalization argument, it was crucial that
we considered operators which are suppressed only by $v_3^4$, yet
here we are dealing with operators suppressed by $v_3^6$.  In
general, this would be a valid objection, but the terms which
interest us here violate the $U(1)_R$-symmetry of the low energy
$SU(2)$ theory, and
a more careful examination of possible contributions to
${\cal H}$ indicates that other possible corrections of this $R$-symmetry
breaking type are
suppressed by further powers of $v_3$.  In particular, these could
arise from contributions to ${\cal H}$ of the form
\beq
{\psi_{12} \psi_{12} \over \psi_{31} \psi_{31}^{\dagger}}.
\eeq
However, such terms do not respect the $U(1)_R$ symmetry of the full
theory.
The integral over $\psi_{12} \psi_{12} \over \psi_{13}^2$ vanishes.
Note that there are possible corrections to the terms involving
$\phi^a \phi^{a\dagger}$, coming from operators such as ${\psi_{12} \over
\psi_3}
{\psi_{12}^{\dagger} \over \psi_3^{\dagger}}$
and our arguments are not powerful enough to determine if these
are or are not renormalized.}.

Treating ${\cal L}_{eff}$ of eqn. \ref{rviolating} as a
perturbation, we want to consider flat directions of the low
energy $SU(2)$ theory, and construct the effective lagrangian in
these flat directions to first order in the perturbation.
We can do this in a manner similar to the treatment of the $F_{\mu \nu}^4$
terms
if we treat the supersymmetry breaking
terms as a spurion, as follows.  Describe the gauge coupling by a chiral
field, $\tau$ (chiral with respect to both $\theta$ and
$\tilde \theta$).  Take the highest component of $\tau$ to
be proportional to $F_{\mu \nu}^4$, i.e.
\beq
\tau = a + {i \over g^2}
+\dots \theta^2 \tilde \theta^2 m^2,
\label{phivev}
\eeq
with
\beq
m^2 = {9 \over 8 \pi^2} {[(F^{[8]}_{\mu \nu})^4+ \dots] \over \vert v_3 \vert^4
(v_3)^{2}}.
\eeq
Then the $R$-symmetry violating $\phi^a \phi^a$
correction to the effective action arises from the coupling of
$\tau$:
\beq
{\cal L}_{eff}= \int d^2 \theta d^2 \tilde \theta \psi^a \psi^a
\tau.
\eeq

In order to determine the terms in the effective action linear
in $m^2$ which violate the $R$-symmetry, we just need to analyze
the possible $\tau$-dependence of the effective action.
The $\tau$-independent terms are just those of the usual
$SU(2)$-theory.  By the same arguments as in \cite{dsnr},
there are no possible $\tau$-dependent terms one can add
to the lagrangian (without covariant
derivatives, i.e. involving less than six derivatives), except for
one which vanishes in the case of constant (lowest component)
$\tau$.  This term is:
\beq
\int d^8 \theta  \ln(\psi) \tau^{\dagger} + {\rm
c.c.}
\eeq
The term is scale invariant.
It is invariant under the $R$
symmetry because under the symmetry
$\ln \psi$ shifts by a constant, and the
remaining integral gives  zero since $\tau^{\dagger}$ is antichiral.
It is the unique term involving an integral over the full superspace with
a non-trivial $\tau$-dependence.  However, it has the wrong $g^2$-dependence
to correspond to Feynman diagrams -- it has too few powers of $g^2$.
So it is not generated in the theory.

So, in fact, it would seem that there are no terms in the
effective theory linear in the symmetry breaking.  This is
a non-renormalization theorem, but it seems too strong.
How are we to
account for the explicit loop corrections in \cite{deg}?  Here,
one must be careful about the use of the equations of motion.  In
the absence of the symmetry-breaking term, $\partial^2 \phi=0$.
As pointed out by \cite{periwal}, the ${\cal N}=2$ action differs from
that computed by the (component) background field method by terms
which vanish by the tree level equations of motion.  Including
the quantum corrections, these terms are sixth order
in derivatives.  In the presence of the
perturbation, however, $\partial^2 \phi =2 m^2 \phi^{\dagger}$.  In this
case, there are additional terms in the effective action.  These
can be worked out using formulas which are conveniently collected
in \cite{rocekmonopoles}.  These authors work out the lagrangian
of eqn. \ref{hform} in components.  Examining their results (eqns.
B.1-B.9 of that paper), there is one term bilinear in the
$\phi$'s:
\beq
{1 \over 2} {\cal H}_{\phi,\bar \phi} \nabla^2 \phi \nabla^2
\phi^{\dagger},
\eeq
Using the equations of motion and the actual form of ${\cal H}$
this yields:
\beq
{1 \over 2 \pi^2} m^2 \phi m^2 \phi^{\dagger}.
\label{usedeom}
\eeq

This is to be compared to the computation of \cite{deg}, where one
studied
\beq
m^2 \langle \phi \phi \rangle + {\rm c.c.}
 = {1 \over 8 \pi^2} {m^2 \over \vert v_{12}
 \vert^2}\partial_{\mu}\phi \partial_{\mu}{\phi} + {\rm c.c.}
 \eeq
which, by the equations of motion, is equal to the expression,
eqn. \ref{usedeom} above, after an integration by parts.

This argument establishes that there are no further
renormalizations of any $R$-symmetry violating terms
in $SU(3)$, with six derivatives.  The
two loop terms
in \cite{deg} are generated by a combination of the four derivative
terms from
integrating out the most massive fields, plus the four
derivative $SU(2)$ terms.  Neither of these are renormalized.

\section{Generalization to $SU(N)$}

In the case of $SU(N)$, we can repeat these arguments. First,
just as for the $F_{\mu \nu}^4$ terms, we can show that certain $F_{\mu \nu}^6$ terms
are not renormalized.
In particular, consider first breaking $SU(N)$ to $SU(N-1)$ by an
expectation value, $v_N$.  We have seen that the four derivative
terms in the one loop effective
action involving $(F^{[N-1]})^4 \over (v_N)^4$
are not renormalized.
This yields the obvious generalization of
the effective action of eqn.
\ref{rviolating}, where the sum now runs over the generators
of the adjoint representation of $SU(N-1)$:
\beq
{\cal L}_{eff}={1 \over 2 \pi^2}{N^2 \over (N-1)^2}
{[(F^{[N-1]}_{\mu \nu})^4+ \dots] \over \vert v_N\vert^4}
({ \vert \phi^a \vert^2 \over \vert v_N \vert
^2}+ {\phi^a \phi^a \over (v_N)^2}+ {\phi^{a\dagger} \phi^{a\dagger} \over
(v_{N}^*)^{2}}).
\label{rviolatingsun}
\eeq
We can again describe the $R$-symmetry
violating part of this perturbation by treating
the highest component of $\tau$, $m^2$, as
a spurion:
\beq
m^2= {1 \over 2 \pi^2}{N^2 \over (N-1)^2}
{[(F^{[N-1]}_{\mu \nu})^4+ \dots] \over \vert v_N
\vert^4 (v_N)^2}
\eeq
We have already established that this term is not
renormalized.  Now we can consider the $SU(N-1)$ theory,
with this interaction as a perturbation. We
consider the breaking of this symmetry to $SU(N-2)$.
The four derivative terms are described
by
\beq
{\cal H}= {1 \over 16 \pi^2}\sum_{i=1}^{N-2}
\ln(\psi_{N-1,i})\ln(\psi_{N-1,i}^{\dagger}).
\eeq
Again, there are no $\tau$-dependent corrections.
Since the perturbation
can be described in terms of a background $\tau$, we see, again,
that up to terms related to the equations of motion, there are no
$\tau$-dependent corrections to the four derivative terms in the
low energy theory, corresponding to the absence of corrections
to six derivative
symmetry violating terms in the full theory.

Now consider the ($\tau$-independent) terms.  We want to consider
these as perturbations in the lower energy $SU(N-2)$ theory.
Rewriting this expression in terms of the Cartan generators, we
obtain:
\beq
{1 \over 2\pi^2}{(N-1)^2 \over
(N-2)^2} {\partial^2 \phi^{[N-2]} \partial^2 \phi^{[N-2] \dagger}
\over \vert v_{N-1} \vert^2} \sum_{i=1}^{N-3} (\phi^{i} \phi^{i} + {\rm c.c.} +
\dots)
\eeq
Here we have kept only terms which are relevant to our analysis,
i.e. those for which the equations of motion will yield factors of
$m^2$.
Again, this can be generalized to an expression invariant under
$SU(N-2)$.  Using the equations of motion, it reduces to the
expression obtained in the (component) background field method.

Further operators can now be obtained by iteration.
Finally, we are left with the operator:
\beq
{1 \over g^2}({g^2\over 2 \pi^2})^{N-1}
\prod_{n=2}^N  {n^2 \over (n-1)^2}
{[(F^{[N-1]}_{\mu \nu})^4+ \dots] \over \vert v_N\vert^4}
({\partial_{\mu} \phi^{[N-2]} \partial^{\mu} \phi^{[N-2]}
\over v_{N}^2 \vert v_{N-1} \vert^2}+ {\rm c.c.})
\eeq
$$
({\partial_{\mu} \phi^{[N-3]} \partial^{\mu} \phi^{[N-3]}
\over v_{N-1}^2 \vert v_{N-2}\vert^2}+ {\rm c.c.})
\dots
({\partial_{\mu} \phi^{[1]} \partial^{\mu} \phi^{[1]}
\over \vert  v_{12}\vert^2}+{\rm c.c.})
$$
In sum, we
see that a set of non-zero terms with a particular symmetry structure
are
obtained by symmetry arguments (up to one
overall coefficient).  They are generated at precisely the
expected order in the coupling, with precisely the values obtained
from explicit component field computations.  Because they are
generated from structures which are not renormalized, these
operators are themselves not renormalized.  So we have exhibited
what we promised:  a set of operators with up to $2N$ derivatives
which are not renormalized.

\section{Finite ${\cal N}=2$ Theories}

In \cite{dsnr}, it was noted that not only are the $F_{\mu \nu}^4$ terms
note renormalized in ${\cal N}=4$ theories, but identical arguments imply
that they are not renormalized in ${\cal N}=2$ theories.  The same
applies to the $2N$ derivative terms we have considered here; the
scale invariance and $R$ symmetries which were necessary
in the ${\cal N}=4$ case also hold in these theories.  All of the
arguments we have given above go through word for word.  The extra
matter multiplets in these theories play a similar role to that of
the hypermultiplets in ${\cal N}=4$ theories.  Like those fields, they
carry charge $+4/3$ under the $U(1)_R$ symmetry.

This represents, then, another large class of theories for which
the coefficients of terms with arbitrarily large numbers of
derivatives can be calculated exactly.

\section{Conclusions}

We have established that in ${\cal N}=4$ Yang-Mills theories, there
is a large set of non-renormalization theorems.
This property, already guessed from the behavior of the matrix
model, is quite remarkable, and one might wonder both about
extensions and possible applications.  Given the complete
agreement of the three graviton scattering amplitude in the case
of the matrix model, we might expect complete agreement
in the field theory for
amplitudes which scale correctly with separation.
Note also that our techniques do not permit study of terms
with more than $2N$ derivatives, which can be generated by
operators with covariant derivatives.

Another question is:  while non-renormalization theorems plus
dualities account for the agreement of the matrix model and
supergravity\cite{seiberg,sen,polchinski}, in what sense do the
agreement of each of the {\it coefficients} of the $2N$
derivative terms between
the matrix model and supergravity constitute independent tests of
the dualities?  Given that the results follow from the structure
of iterations of the one loop action, one suspects that the answer
is that they do not.

These observations should also have implications for the understanding
of the AdS/CFT correspondence.  It should be possible to generalize
the analyses of \cite{douglastaylor,das} for scattering of two
D3-branes to N D3-branes in an AdS background.  This is currently under
investigation.
Finally, one can ask what sorts of non-perturbative information
can be extracted from the theory using these results, and whether
they are applicable to other non-trivial field theories.
What other interesting facts may be
gleaned about these theories, as well as more complete answers to
the questions raised above, are all subjects worthy of further
study.

\noindent {\bf Acknowledgements:}

\noindent
We are grateful to Tom Banks, Nathan Seiberg, Steve Shenker and Scott Thomas
for critical comments and suggestions. Correspondence
with Yuri Shirman and Rikard von Unge
was helpful in clarifying many issues.  Comments
by J. Polchinski, S. Giddings, and others at the ITP
seminar of M.D. helped to clarify possible connections
to AdS.   This work supported in
part by the U.S. Department of Energy.


\end{document}